\documentclass[sigconf]{acmart}

\AtBeginDocument{%
  \providecommand\BibTeX{{%
    \normalfont B\kern-0.5em{\scshape i\kern-0.25em b}\kern-0.8em\TeX}}}

\copyrightyear{2021} 
\acmYear{2021} 
\setcopyright{acmcopyright}\acmConference[DIS '21]{Designing Interactive Systems Conference 2021}{June 28-July 2, 2021}{Virtual Event, USA}
\acmBooktitle{Designing Interactive Systems Conference 2021 (DIS '21), June 28-July 2, 2021, Virtual Event, USA}
\acmPrice{15.00}
\acmDOI{10.1145/3461778.3462004}
\acmISBN{978-1-4503-8476-6/21/06}

\usepackage{balance}       
\usepackage{graphics}      
\usepackage{hyperref}
\usepackage{color}
\usepackage{booktabs}
\usepackage{textcomp}

\usepackage{tabularx}
\usepackage{multirow}
\usepackage{array}
\usepackage{amsmath}
\usepackage{stfloats}
\usepackage{graphicx}
\usepackage{amsthm}
\usepackage{listings}
\usepackage{caption} 
\usepackage[linesnumbered]{algorithm2e}
\usepackage{soul}

\newcommand*{\eg}{\textit{e.g.},\xspace}
\newcommand*{\vs}{\textit{v.s.}\xspace}
\begin{document}

\title{Understanding the Design Space of Mouth Microgestures}


\author{Victor Chen}
\email{vchen36@stanford.edu}
\affiliation{%
  \institution{Department of Computer Science, Stanford University}
  \city{Stanford}
  \state{CA}
  \country{USA}
  \postcode{94305}
}

\author{Xuhai Xu}
\email{xuhaixu@uw.edu}
\affiliation{%
  \institution{Information School, University of Washington}
  \city{Seattle}
   \state{WA}
   \country{USA}
   \postcode{98105}
}

\author{Richard Li}
\email{lichard@cs.washington.edu}
\affiliation{%
  \institution{Paul G. Allen School of Computer Science \& Engineering, University of Washington}
   \city{Seattle}
   \state{WA}
   \country{USA}
   \postcode{98105}
}

\author{Yuanchun Shi}
\email{shiyc@tsinghua.edu.cn}
\affiliation{%
  \institution{Department of Computer Science and Technology, Key Laboratory of Pervasive Computing, Ministry of Education, Tsinghua University}
   \city{Beijing}
   \country{China}
   \postcode{100084}
}
\author{Shwetak Patel}
\email{shwetak@cs.washington.edu}
\affiliation{%
  \institution{Paul G. Allen School of Computer Science \& Engineering, University of Washington}
   \city{Seattle}
   \state{WA}
   \country{USA}
   \postcode{98105}
}

\author{Yuntao Wang}
 \authornote{This is the corresponding author.}
\email{yuntaowang@tsinghua.edu.cn}
\affiliation{%
  \institution{Department of Computer Science and Technology, Key Laboratory of Pervasive Computing, Ministry of Education, Tsinghua University}
   \city{Beijing}
   \country{China}
   \postcode{100084}
}

\renewcommand{\shortauthors}{Chen et al.}

\begin{abstract}
As wearable devices move toward the face (i.e. smart earbuds, glasses), there is an increasing need to facilitate intuitive interactions with these devices. Current sensing techniques can already detect many mouth-based gestures; however, users’ preferences of these gestures are not fully understood. In this paper, we investigate the design space and usability of mouth-based microgestures. We first conducted brainstorming sessions (N=16) and compiled an extensive set of 86 user-defined gestures. Then, with an online survey (N=50), we assessed the physical and mental demand of our gesture set and identified a subset of 14 gestures that can be performed easily and naturally. Finally, we conducted a remote Wizard-of-Oz usability study (N=11) mapping gestures to various daily smartphone operations under a sitting and walking context. From these studies, we develop a taxonomy for mouth gestures, finalize a practical gesture set for common applications, and provide design guidelines for future mouth-based gesture interactions.


\end{abstract}



\maketitle

\section{Introduction}
\label{sec:intro}


Since the early 1990s, researchers have been investigating using the mouth as an eyes-free and hands-free input channel to facilitate human-computer interaction~\cite{salem1997isometric}.
Nowadays, with the advances of electronic technology, wearable devices such as earbuds and head-mounted displays are becoming increasingly ubiquitous and providing new sensing techniques around the face and the mouth.
Therefore, there has been emerging research on mouth-related gestures recently, such as teeth clicking~\cite{Ashbrook2016,10.1145/3290607.3312925,Xu}, humming~\cite{jylha2011sonic}, or chewing~\cite{Cascon2019}. These gestures are usually subtle, requiring little effort from users, enabling eyes- and hands-free interactions, and having a tendency to be more socially acceptable.

However, in spite of the rich prior work, there is still a lack of an overall understanding of the design space of \textit{mouth microgestures}, which we define as \textit{any deliberate action or movement involving any part of the mouth with the purpose of controlling some device}.
It is also unclear whether users would prefer a certain set of gestures over others.
Previous studies in this space were conducted independently, making it difficult to compare their findings.
Moreover, different sensing modalities might affect or bias the user experience, making the comparison even more difficult.
To the best of our knowledge, there is no prior work fully exploring the design space of mouth microgestures, including users' preference in the space. We seek to provide a foundation of this novel design space for interacting with next generation mobile and wearable devices.
With an unexplored interaction space such as this, it is important that we first understand what users envision as an ideal gesture, free of the constraints of sensing technology. This knowledge can then help guide future researchers in designing usable systems that people will actively adopt.

In this paper, we conducted a series of user studies to explore and evaluate the design space of mouth microgestures. Note, we will use the terms ``microgesture'' and ``gesture'' interchangeably. 
First, we organized four remote brainstorming sessions obtaining a set of 86 mouth microgestures. From this, we derived a taxonomy of these gestures based on the mouth organs used in the gesture as well as the primary form of how the gesture is represented. Gestures that use the tongue as the main active organ were proposed the most, followed by those of the outer mouth parts (e.g. lips). Next, we conducted an online survey comprising pairwise comparisons of the proposed user-defined microgestures in terms of physical and mental workload. 
After forming a ranking of the microgestures from the results, we select a preferred subset of 20 gestures from the original gesture set that contains both the least physically and mentally demanding microgestures. Consequently, we conduct a remote Wizard-of-Oz user study to map the subset of microgestures to real-life tasks users would do under two contexts: sitting and walking. We use the participants' agreement on microgestures across tasks to form a practical mouth microgesture set and describe quantitative and qualitative insights from the results of our user studies.

This paper's contributions are threefold:
\begin{enumerate}
    \item We qualitatively examine various characteristics of user-defined mouth gestures and develop a taxonomy of mouth gestures for understanding the design space in regards to the mouth organs used, modality, and expressive power.
    \item Through two user studies, we obtain a practical mouth gesture set for daily tasks for common applications under real-life contexts.
    \item With insights about users' preferences, we provide a set of design implications and recommendations for how HCI researchers and designers can design new and usable mouth gesture interfaces.
\end{enumerate}

\section{Related Work}
\label{sec:related}
We first review the existing systems of mouth-related gestures.
We also summarize user-defined gesture design, which is adopted by our first user study.

\subsection{Mouth-based gesture interactions }

A large body of literature has explored the 
idea of mouth-based gestures.
Many of these systems have been used as assistive technologies, providing an effective alternative mode of interaction for those with diseases or injuries that limit many motor skills~\cite{kim2013tongue}.
Indeed, mouth-based gesture controls could be used to mitigate permanent, temporary, and situational impairments, suggesting an inclusive design framework that could be applied to all kinds of users.
Included in these impairments is the visual attention needed to operate many current consumer devices~\cite{oakley2007designing}, with mouth-based gestures potentially alleviating the related cognitive load.
In addition to convenience, mouth-based gesture controls can be very subtle due to the fine control humans have over their mouths, similar in granularity to hand dexterity~\cite{gentilucci2001grasp}, and tongue movements being containable within a closed mouth.
As a result, mouth-based gesture controls can have potential advantages as an alternative to speech recognition in situations where users are in either a very quiet environment or a very noisy environment, where speaking aloud is not appropriate or difficult to sense~\cite{denby2010silent}.


Prior work has focused on developing technical sensing systems to test the feasibility for detecting certain gestures. 
While some works do evaluate qualitative aspects of their gesture interface from a user experience angle~\cite{Nguyen-TYTH, Xu, goel2015tongue}, most mainly assess the performance of their system itself or discuss the user preference as a secondary point. 
Due to the constraints of sensors, they often target a specific sub-part of the mouth that is propitious for their technical approach. 
In addition to having possible bias in users' opinions from the physical system, these works, which focus on different parts of the mouth, are not easily comparable in their results.
Proximity makes it so parts of the mouth often interact with each other which lends itself to treat the mouth as a whole as an interface. By considering it this way, we also hope to gain common insights that are applicable to an interaction regardless of which part of the mouth is used.

Although there is a gap in this design knowledge, we review the papers proposing technical systems for sub-parts of the mouth and their insights to serve as motivation that a mouth-based gesture interface is a practical and valuable design space.

\subsubsection{Tongue interactions}
The tongue is capable of a high degree of expression and dexterity \cite{Niu2019}; much prior work has exploited the benefits of these fine motor capabilities to create tongue-based interfaces. \textit{TYTH} from Nguyen et al. \cite{Nguyen-TYTH} show that the tongue can be used to accurately tap different areas of teeth to enable typing like a keyboard;
users approved of the interaction but did not support the physical form factor as much.
Other studies involve equipping the tongue with a magnet for precise tracking to control an interface \cite{sahni2014tongue}. Tongueboard \cite{Li-TongueBoard} detects input from the tongue through an oral retainer with capacitive touch sensors. Less invasive techniques have also been explored for tongue input, such as using RGB cameras for tracking \cite{Niu2019} or attaching a pressure-based interface to the outer cheek \cite{cheng2014tip}. 

\subsubsection{Teeth interactions}

Prior literature has explored the feasibility of using tooth clicking as input, particularly in the accessibility field. Zhong et al. and Kuzume et al. both used in-ear bone conduction microphones to detect the occurrence of a tooth click~\cite{zhong2007osteoconduct,kuzume2012evaluation}. Bitey~\cite{Ashbrook2016} expands upon this work to allow for distinguishing different pairs of teeth clicking, and Byte.it~\cite{10.1145/3290607.3312925} demonstrates that the interaction technique can be implemented with other commodity sensors like an accelerometer or gyroscope. Additionally, Xu et al. proposed a system of clench interactions that differentiate different degrees of force when biting down 
and found that users appreciated the clench interaction as a hands-free technique
~\cite{Xu}.

\subsubsection{Face-related interactions}

Facial movements and expressions are a natural, common occurrence in everyday human behavior. Recently, researchers have explored systems that can not only recognize these facial muscle movements but also leverage them to serve as a way to directly manipulate interfaces~\cite{lyons2004facial,10.1145/3313831.3376810,10.1145/3313831.3376836}. Among the many areas of the face, the various parts of the mouth, both internal and external, and the ways they move and interact with each other are particular points of interest for interaction design. 
Beyond simple facial recognition, camera-based techniques have been used to track lips to perform lip reading \cite{shillingford2018large, sun2018lip}. Movements of the eyes, eyebrows, and mouth have been shown to be recognizable with electrooculography (EOG) and electromyography (EMG) sensors applied to the face~\cite{paul2013smart,nakao2018make}. Tongue-in-Cheek allows for gestures using the tongue, cheek, or jaw to be detected in a wireless, non-invasive manner for directional input~\cite{goel2015tongue}. 
Their user study showed that their system was preferred, but the target user group was those with neuromuscular conditions and not the general population.
A similarly non-contact solution uses proximity sensors to enable continuous tracking of the cheeks and jaw from a virtual reality headset~\cite{li2018buccal}. Research involving outer ear interfaces (OEI) have also demonstrated that deformations of the ear canal can be sensed to detect facial movements and expressions as a form of input~\cite{Matthies2017,Amesaka2019,Ando2017}.


\subsection{User-defined gesture design}

When new systems using gesture interfaces are developed, the design of the gestures is often constrained by technical feasibility or the knowledge of those implementing the system. Participatory design is a well-studied approach that integrates users into the decision-making and design process, and this method is valuable for designing gesture interfaces as well~\cite{schuler1993participatory}. Gesture elicitation studies were proposed by Wobbrock et al. for interactive surface computing~\cite{wobbrock2005maximizing}. These studies follow a procedure where the participant is shown the effect of an action, called the \textit{referent}, and asked to provide the \textit{sign}, the gesture that would produce the \textit{referent}. Compared to gestures designed by experts, user-defined gestures have been found to be more intuitive to learn and easily memorable for end-users~\cite{morris2010understanding,nacenta2013memorability}. A user elicitation technique also produces gestures covering a much wider scope as well as being more preferred than those human-computer interaction experts could generate~\cite{wobbrock-user-defined}. Over the years, gesture elicitation studies have been applied to a wide variety of gesture interactions~\cite{villarreal2020systematic}, such as those using the hand~\cite{chan2016user}, foot~\cite{fukahori2015exploring}, head movements~\cite{yan2018headgesture}, face~\cite{lai2020understanding}, and fingers~\cite{gheran2018gestures,fariman2016designing}. 

\smallskip
In our work, we take a similar approach to include users' input for the design of gestures specifically focused around the mouth. Inspired by prior work using a framed guessability methodology~\cite{cafaro2018framed}, we employ both open elicitation, where users are unconstrained when proposing gestures, and closed elicitation, where users can only select from a smaller, focused set of gestures~\cite{villarreal2020systematic}. 
\section{Study 1: Brainstorming Mouth Microgestures}
\label{sec:study1}

To better understand the design space of mouth microgestures, we first needed to gather a detailed list of possible microgestures that can be performed by users.
In this study, we invited 16 participants across four sessions to  brainstorm and design mouth microgestures.

\subsection{Participants}

We recruited 16 participants through mailing lists and online communities.
Eleven were male and five were female, with ages ranging from 20-34 (mean=24, stdev=3.96).
Eleven out of sixteen participants (69\%) came from a technical or engineering background.
Others worked in areas including design, medicine, and education.
All but two had experience with wearable devices, and four self-reported never having used gesture control for a device before.

\subsection{Procedure}

Participants were placed in groups of four along with one of the researchers as a moderator. Groups collaborated remotely over a video conference call. Each brainstorming session, which lasted for about one hour, began with a short icebreaker question and introductions so that participants could familiarize themselves with each other. The moderator then described the purpose and procedure for the brainstorming session, as well as the definition of a mouth microgesture. For this study, we kept the goal of the session open-ended and asked the participants to simply brainstorm as many microgestures as they could, without considering the sensing feasibility; however, we did not describe specific applications or tasks from which to base their ideas. This choice in procedure was to keep the participants' focus on the physical nature of using the different parts of the mouth to perform microgestures. With more general guidelines, participants would not need to concern themselves with other aspects of their ideas, such as whether it is plausible to detect with current technology or if it is easy to perform. The only limitation we imposed on the design was that microgestures involving spoken words should be avoided, since speech commands as an interface follow and carry different interaction principles than microgestures do. 

To record their ideas, participants used an online collaborative whiteboard tool called Stormboard~\footnote{https://www.stormboard.com/} to write down their proposed microgestures on virtual sticky notes. An individual brainstorming period was conducted first for 10 minutes where participants worked separately to think of ideas. Next, the participants were brought back together and took turns discussing their ideas, bringing all of their sticky notes into a shared workspace for the whole group to view. After sharing, participants were asked to spend the rest of the time working together to create new ideas, adding to those they thought of individually. The moderator presented questions to spark new lines of thought whenever the group had trouble brainstorming new ideas.

\begin{figure*}[]
    \begin{minipage}[t]{0.45\textwidth}
        \centering
        \includegraphics[width=\linewidth]{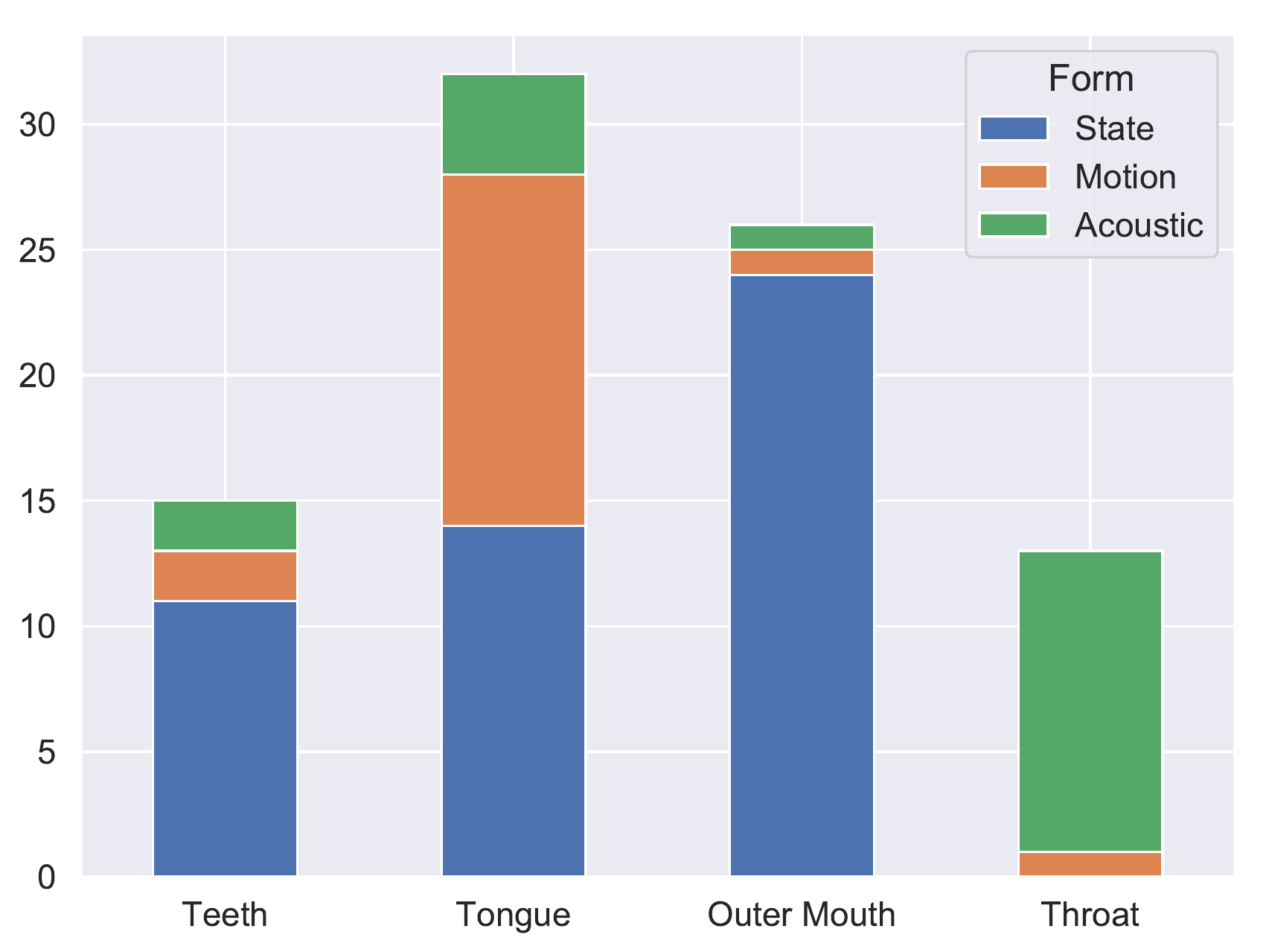}
        \makebox[\linewidth]{\small (a)}%
        \label{stackedbar}
    \end{minipage}
    \begin{minipage}[t]{0.54\textwidth}
        \centering
        \includegraphics[width=\linewidth]{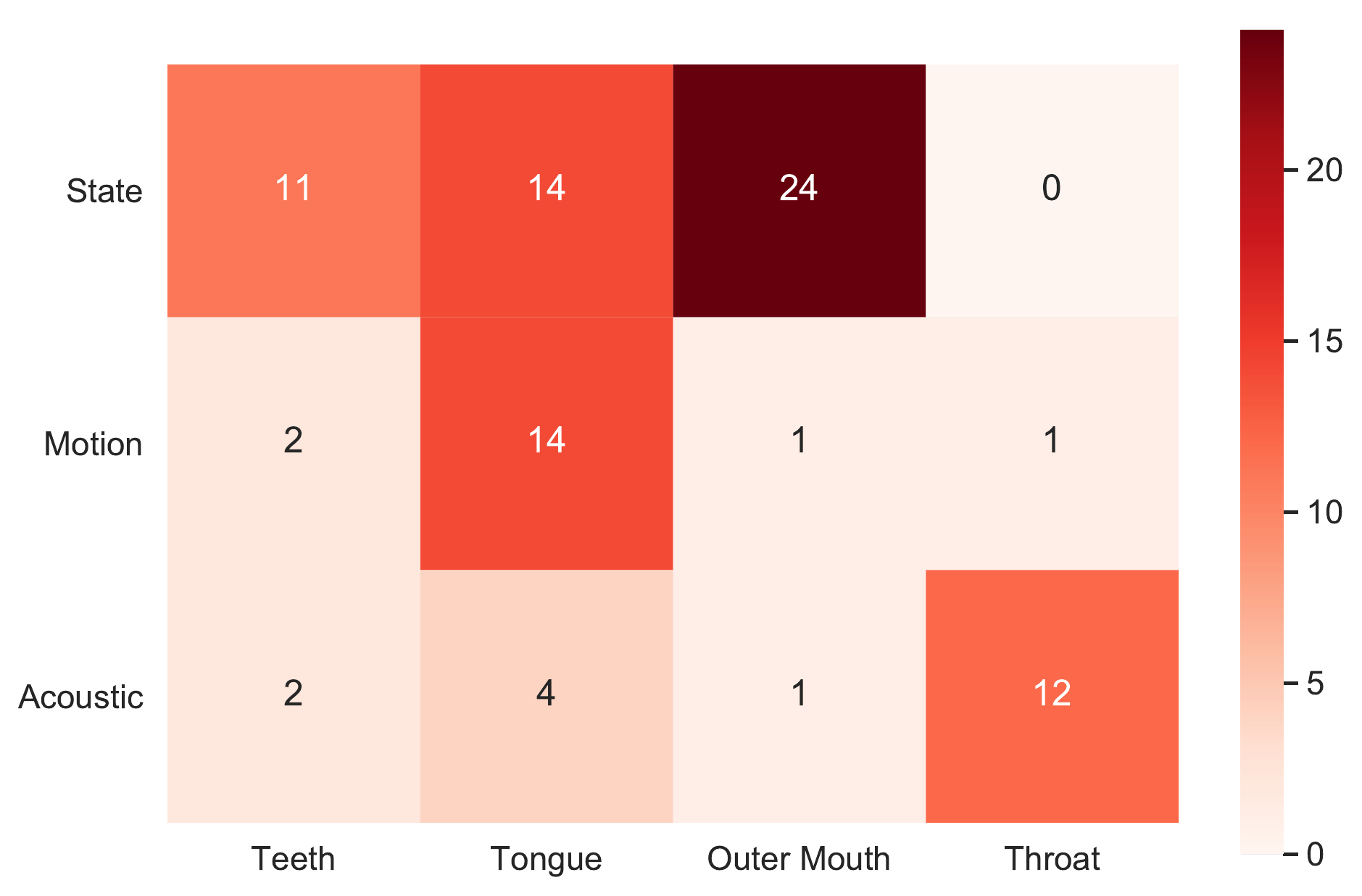}

        \makebox[\linewidth]{\small (b)}%
        \label{heatmap}
    \end{minipage}

    \caption[]{Chart (a) shows the distribution of the user-defined gesture set by the counts of gestures with a particular main \textit{actor} organ. Each bar also shows the composition of the category based on the form of the gesture. Heatmap (b) shows the count density of unique brainstormed ideas according to the mouth organ and form.}
    \label{fig:gesture_dist}
\end{figure*}

\subsection{Results}
\label{sub:study1:results}
Across the four brainstorming sessions, a total of 104 unique ideas were proposed. We found that many of them were simply variants of other microgestures, such as repeated actions (tapping two times or three times) or duration-dependent actions (holding a pose for one second vs two seconds).
In this first study, we were interested in compiling a set of fundamental, unique units of gesture that could be compared fairly in the following study described in Section \ref{sec:study2}.
For example, it would be unfair to compare tapping the tongue to the roof of the mouth twice versus once based on physical demand. 
Therefore, we merged these gestures if they only differed on repetition and time. We additionally filtered out gestures that did not fit our definition of mouth microgesture or could not be performed by the general population, leaving us with 86 unique mouth microgestures that are distinct in the physical motion of the mouth organ(s).

We create a taxonomy of our full gesture set along two axes. The first is based on the parts of the mouth used to perform the microgesture and the way they interact with each other. We refer to this as the \textit{actor-receiver} pattern. To perform a mouth microgesture, there is often one mouth organ acting upon another. The \textit{actor} refers to the primary organ that is moving or controlling the gesture; the \textit{receiver} is the organ that is receiving the motion or action from the \textit{actor}. For example, in the gesture \textit{bite tongue with left side of teeth}, the \textit{actor} organ would be the teeth, since the main motion of the gesture is the biting down action, and the \textit{receiver} organ would be the tongue, since the tongue is being acted upon by the teeth. If the gesture only involves one part of the mouth, then that part serves as both the \textit{actor} and \textit{receiver}.
Using this pattern, we define four categories of one axis of our taxonomy by the \textit{actor} in the relationship: \textit{teeth}, \textit{tongue}, \textit{outer mouth}, and \textit{throat}. Note that the outer mouth refers to the different external areas/muscles of the mouth like the lips, cheeks, and jaw.

The second axis is determined by the form or modality with which the microgesture is executed. We characterize our gesture set into three of these categories: \textit{state}, \textit{motion}, and \textit{acoustic}. Although all microgestures contain some degree of motion, \textit{state} describes microgestures where the goal of the movement is to reach some condition or position for however brief a period. The gesture \textit{bite down on tongue with front teeth} falls in this category, because the intent of the gesture is to arrive at a state where the tongue is being bitten down on. For the \textit{motion} category, the movement of the mouth organ itself, between start and finish of execution, defines the microgesture. An example would be \textit{slide tongue forward on roof of mouth}; the sliding motion of the tongue is what characterizes this gesture. The last category, \textit{acoustic}, describes gestures that produce a unique sound from the movement of mouth organs, such as \textit{clicking the tongue}. 

Figure \ref{fig:gesture_dist} shows the distribution of the proposed gestures across two axes.
Tongue and outer mouth gestures, each with a similar total number of gestures (29 and 26, respectively), made up the majority of the brainstormed gestures. Most of the outer mouth gestures are \textit{state} gestures, while for tongue gestures, \textit{state} and \textit{motion} gestures almost evenly form the majority. When examining the \textit{state}, \textit{motion}, and \textit{acoustic} categories, we also observe that the \textit{state} group is the largest (49).

\section{Study 2: User-defined Gesture Evaluation}
\label{sec:study2}

With a large set of user-defined microgestures, we conducted a second user study to analyze key features of the gestures that can influence user preference. Motor fatigue and cognitive fatigue are two major concerns when developing a new interaction technique, so we focused on examining and comparing the physical and mental workload of the proposed microgestures. 

\subsection{Categorization of proposed microgestures}

To understand the physical and mental demand of different microgestures, a direct comparison between any arbitrary microgestures would not be valid. Some microgestures may have a form of motion that is more expressive or allows for a more intuitive experience for complex interface operations but at the cost of being more physically demanding than other simple microgestures. Moreover, certain microgestures have an intrinsic, analogous microgesture because of the way direction or location plays a role in the action. For example, the gesture \textit{slide tongue to the right along the bottom lip} would have a corresponding microgesture but sliding toward the left instead. We grouped these such microgestures into pairs for the purpose of this comparison study.

In order to fairly compare microgestures against each other, we created three divisions of microgestures based on their information transfer bandwidth: zero-, one-, and multi-bit
. Zero-bit microgestures can be described as those that are standalone and represent a single state of interaction. One-bit microgestures are those that function like a pair, as described earlier; their expressive power can represent two different states. Multi-bit microgestures are fluid and variable in how they are performed; they are the most expressive gestures and produce at least two distinct effects in interaction. This taxonomy enables a fair comparison of microgestures \textit{within} each category. Grouping this way culminates to 19 zero-bit, 28 one-bit, and 13 multi-bit microgestures. 

\subsection{Procedure}

Even by consolidating the initial gesture set into the aforementioned categories, there are still too many microgestures in each category to allow for comparing physical and mental demand for every combination of microgesture ($C^{2}_{19}=171,C^{2}_{28}=378,C^{2}_{13}=78$).
We overcome this problem by using a pairwise ranking scheme based on the Crowd-BT algorithm~\cite{chen2013pairwise}, which establishes a high quality ranking of microgestures from multiple users while allowing each user to only compare a limited number of comparisons.

We implemented a survey as a web application participants could visit to evaluate pairs of microgestures~\cite{athalye2016gavel}. Participants considered a subset (a third) of the microgestures from each of the three categories (zero-, one-, and multi-bit) for both physical and mental demand, resulting in six phases of the survey. Before each phase, the participants were shown the NASA TLX description for either physical or mental demand~\cite{hart1988development}. Once a phase was begun, at any one time, participants would be presented with text descriptions of two microgestures from the same category as well as a question asking which of the two was less demanding in regard to either physical or mental demand. Four instructional manipulation checks were administered throughout the survey to check for participants' attention~\cite{oppenheimer2009instructional}. We recruited 50 participants using Amazon Mechanical Turk with the requirement that they must have completed at least 5000 tasks previously and have a task approval rating of at least 95\%.

\begin{table*}[]

\small
\begin{tabular}{crl}
\specialrule{1pt}{1pt}{1pt}
\hline
\textbf{Category}  & \textbf{Mean Score}   & \textbf{Gesture Description}  \\
\hline
\multirow{10}{*}{Zero-bit}   & 0.719  & Grind teeth together gently \\
                             & 0.617  & Tap tongue to roof of mouth \\
                             & 0.453  & Stick tongue out \\
                             & 0.423  & Curl tongue upwards \\
                             & 0.400  & Bite down on tongue with front teeth \\
\cline{2-3}
                             & -0.364  & Smack lips \\
                             & -0.471  & Make a short "snoring" sound \\
                             & -0.523  & Trill the tongue or roll the "R" sound\\
                             & -0.852  & Blow a raspberry \\
                             & -1.447  & Chattering teeth \\
\hline
\multirow{12}{*}{One-bit}    & 1.373  & Open or close mouth \\
                             & 1.155  & Smile or frown \\
                             & 1.123  & Slide tongue to the left or right along top lip \\
                             & 0.815  & Slide tongue to the left or right along inner surface of top teeth \\
                             & 0.774  & Blow air out or suck air in \\
                             & 0.609  & Tap tongue to the front, inner surface of top or bottom teeth\\
\cline{2-3}
                             & -0.564  & Slide tongue to the left or right on top of bottom teeth \\
                             & -0.594  & Bite tongue with left or right side of teeth \\
                             & -0.637  & Bite left or right inner cheeks \\
                             & -0.788  & Tuck in/cover up the top or bottom lip with the other lip \\
                             & -1.088  & Click tongue with mouth open to the left or right \\
                             & -1.137  & Shift jaw forward or backward \\
\hline
\multirow{6}{*}{Multi-bit}   & 1.048  & Touch different areas of inner surface of bottom teeth with tongue \\
                             & 0.938  & Touch different areas of inner surface of top teeth with tongue \\
                             & 0.491  & Open mouth to different degrees (open a little bit, open wide, etc.) \\
\cline{2-3}
                             & -0.369  & Move tongue around like pointer/mouse cursor  \\
                             & -1.160  & Touch different areas of outer surface of bottom teeth with tongue \\
                             & -1.196  & Hold tongue in air at different angles/heights \\
\hline
\specialrule{1pt}{1pt}{1pt}
\end{tabular}
\caption{Average of physical and mental demand scores of microgestures in the top and bottom 20\% for each n-bit category. Higher scores indicate that the gesture was less physically and mentally demanding.}
\label{tab:physicalmentalscores}

\end{table*}

\subsection{Results}
\label{subsub:study2:results}
\begin{figure*}[]
    \begin{minipage}[t]{0.48\textwidth}
        \vspace{0pt}
        \centering
            \includegraphics[width=\linewidth]{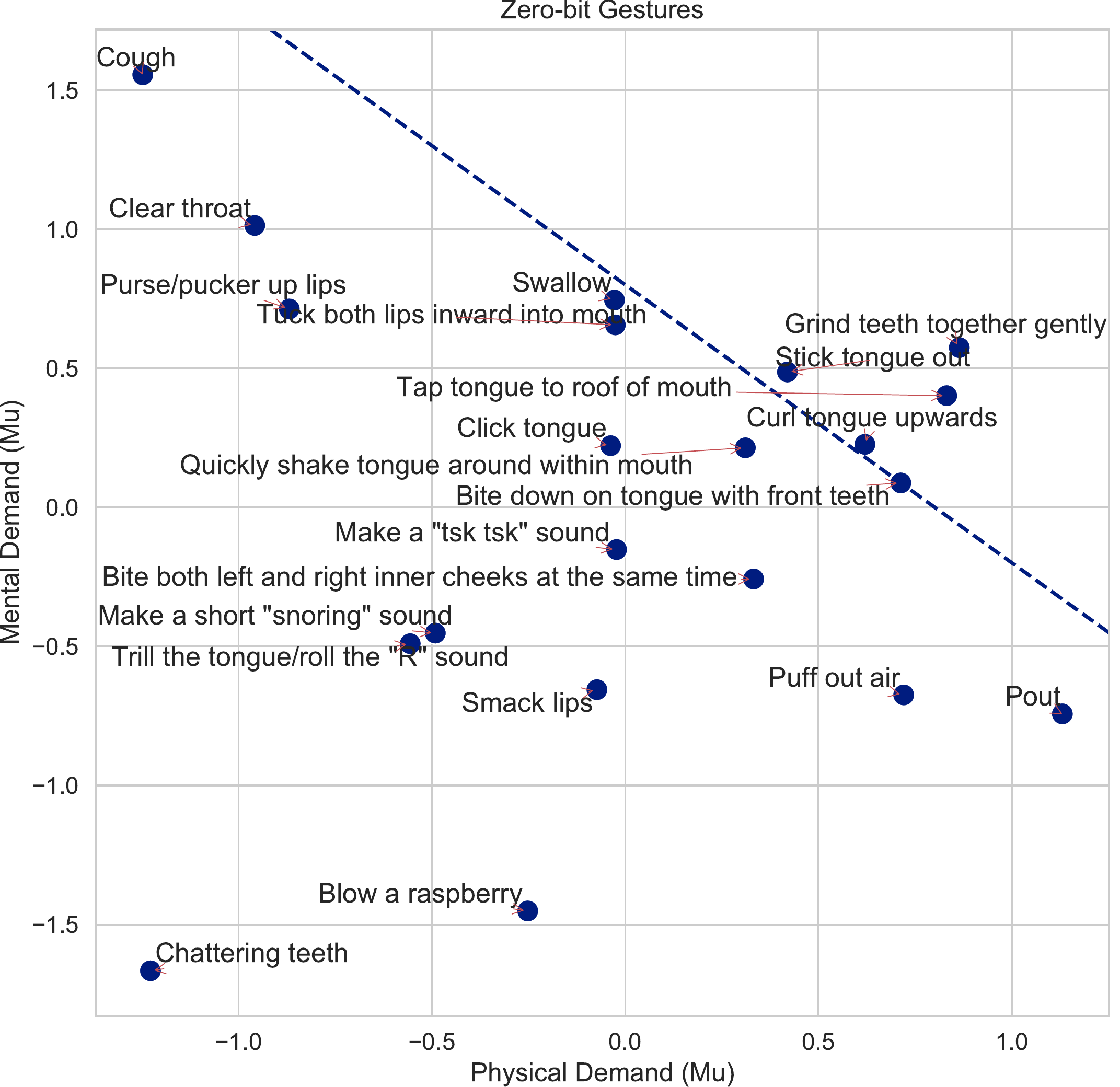}
            \label{zerobit_plot}
        \end{minipage}%
        \hfill
    \begin{minipage}[t]{0.48\textwidth}
        \vspace{0pt}
        \centering
            \includegraphics[width=\linewidth]{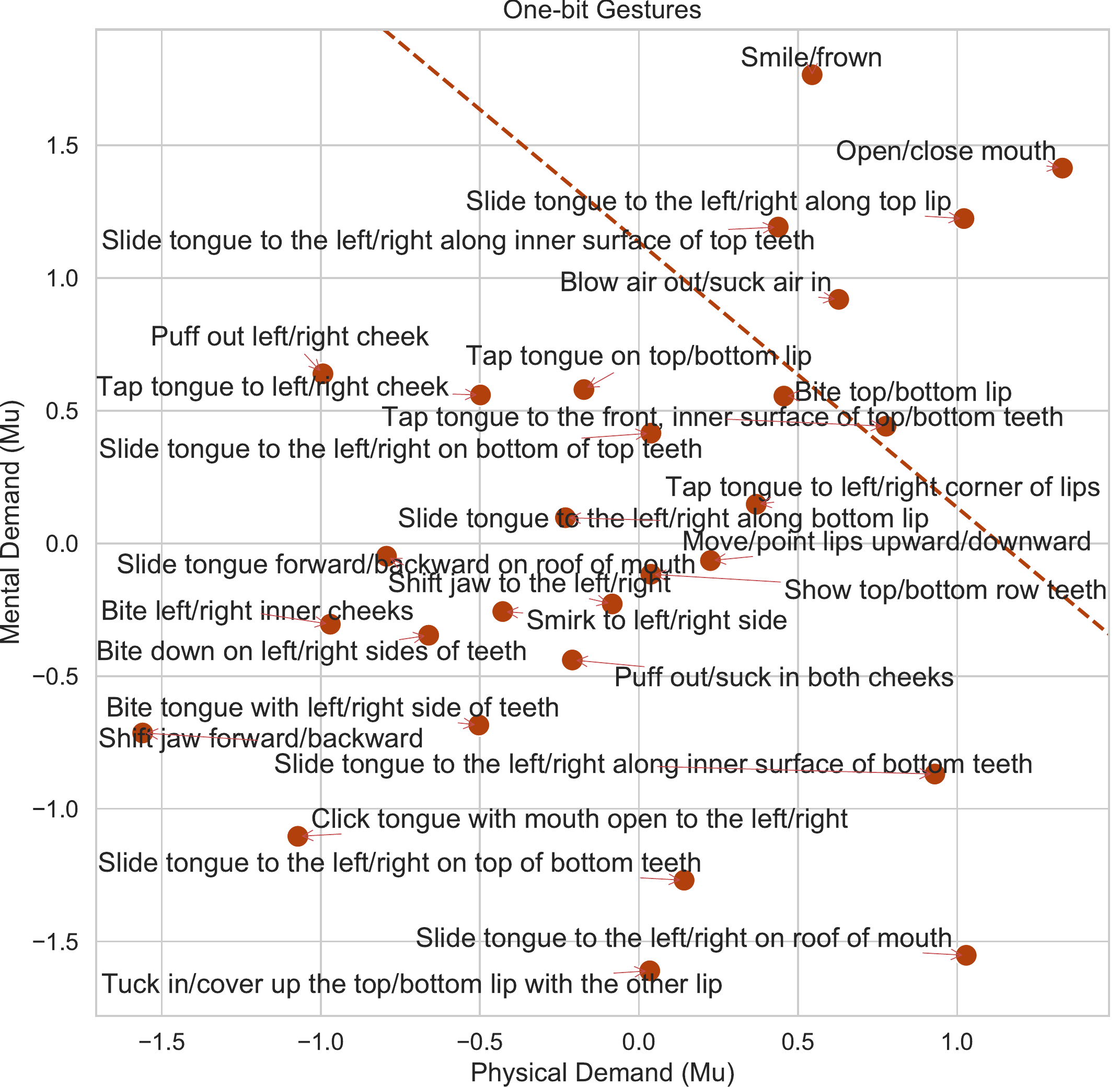}
            \label{onebit_plot}
    \end{minipage}
    \newline
    \begin{minipage}[t]{0.48\textwidth}
        \vspace{0pt}
        \centering
        \includegraphics[width=\linewidth]{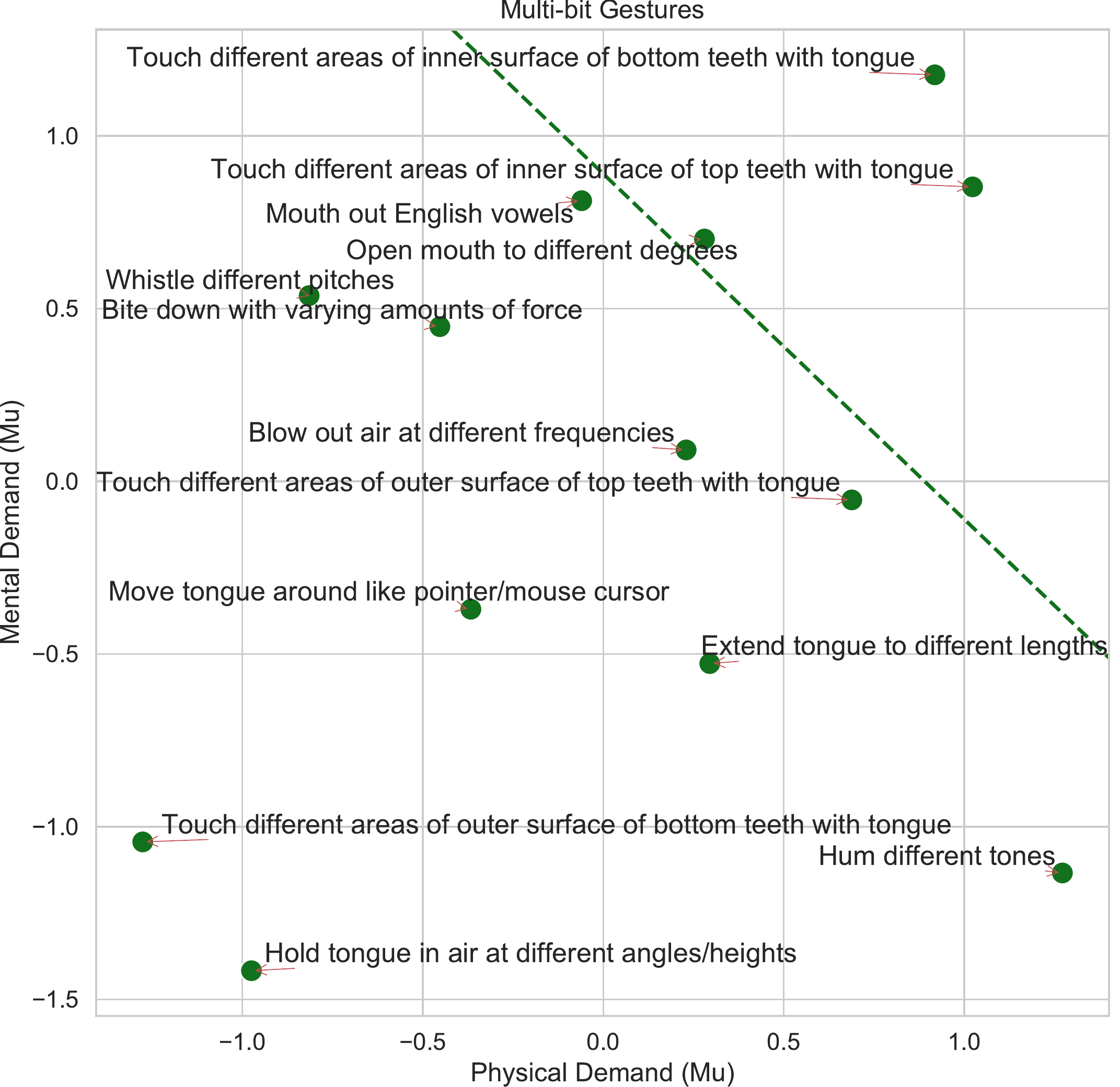}
        \label{multibit_plot}
    \end{minipage}%
        \caption{Physical demand versus mental demand of microgestures from each n-bit category. The higher the mu score is, the less demanding (physical or mental) the gesture was rated. Gestures that are both the least physically and mentally demanding are located in the upper right corner of each visualization. Gestures that are above the dotted line are those in the top 20\% of their category after averaging their physical and mental demand scores.}
        \label{fig:physical_mental_plot}
    
\end{figure*}

In total, we collected 2000 comparisons across all microgesture categories, split evenly between physical or mental demand. In detail, for each type of demand criteria, the 19 zero-bit microgestures had 300 comparisons; the 28 one-bit microgestures had 450 comparisons; the 13 multi-bit microgestures had 250 comparisons. 
We obtained a ranking and raw scores for microgestures in each category based on the inferred quality, calculated by the Crowd-BT algorithm, on both physical and mental demand. 

To help visualize our results, we plotted the physical and mental demand scores against each other. As shown in Figure~\ref{fig:physical_mental_plot}, for the one- and multi-bit categories, there is a clear subset of microgestures in the upper right quadrant that are favored for being both the least physically and mentally demanding in their respective groups. Interestingly, the zero-bit microgestures did not have such a clear consensus between both criteria; some that are less physically demanding end up being more mentally demanding and vice versa. The extreme of this is noticeable with the microgestures \textit{cough}, \textit{clear throat}, and \textit{pucker lips}. These were ranked to be the least mentally demanding but also the most physically demanding of the zero-bit gestures. A similar observation can be found in the other categories with the microgestures \textit{slide tongue to the left/right along roof of mouth} (one-bit) and \textit{hum different tones} (multi-bit), which are among the least physically demanding but also among the most mentally demanding gestures in their categories. Although microgestures like these may not be ideal in one criterion, we believe that since they excel in the other criterion, they may still be usable in certain applications or scenarios. 

For further analysis and later evaluation, we took a subset of microgestures from each category that are both less physically and mentally demanding. To make this selection we averaged the physical and mental demand score of each gesture and then selected the top 20\% from each category. This selection process resulted in 14 remaining gestures as seen in Table \ref{tab:gesture_list} (one-bit gestures are separated into two zero-bit gestures, resulting in 20 gestures). Using the actor-receiver mouth organ taxonomy as described in Section \ref{sub:study1:results}, we noticed that 8 out of the 14 top gestures use the tongue as the actor organ; 3 use the outer mouth; 2 use the teeth; and 1 uses the throat. This is not to say though that microgestures using the tongue are preferred over others. From Figure~\ref{fig:gesture_dist}, we see that tongue microgestures constitute a large portion of the total gesture set. Of the bottom 20\% microgestures, 7 out of 14 are also tongue microgestures. When we take a closer look at the spatial relationships of the motion as well as between the actor and receiver organs, we note a few observations that may point to why some microgestures were favored over others. 

We believe that microgestures that lack sufficient feedback (through proprioception) and require the user to maintain an unnatural state for long periods perform poorer than others. Two of the multi-bit tongue gestures in the bottom 20\% were \textit{move tongue around like a pointer/cursor} and \textit{hold tongue in air at different angles/heights}. With the tongue as the actor organ, these microgestures do not allow the user to easily keep track of the spatial position of the tongue and also require the user to maintain a stretched out tongue which is unnatural. The microgesture \textit{Open mouth at different widths}, which was one of the preferred gestures, at first glance seems to also lack sufficient feedback, but we believe that the motion of opening and closing the mouth is universal and natural enough that the spatial mental model of opening the mouth more or less leads to less physical and mental demand. Indeed, humans have been shown to have control over opening and closing their mouths with similar granularity as hand dexterity~\cite{gentilucci2001grasp}. In the rankings of the zero-bit microgestures as well, we see that \textit{trill the tongue} and \textit{blow raspberry} are not preferred since they are complex actions that need to be held for a period. 

The results of the ranking also suggest that microgestures involving fewer moving organs are less demanding than those that need the user to move multiple parts of the mouth. When considering one-bit microgestures, we see that \textit{bite tongue with left or right side of teeth}, \textit{bite left or right inner cheeks}, and \textit{click tongue with mouth open to the left or right}, which are in the bottom 20\%, all need the user to actively move two different parts of the mouth. 
Whereas with the top 20\% of gestures, even if the actor and receiver organs are different, the receiver is stationary (e.g \textit{slide tongue to the left or right along inner surface of top teeth}).

\section{Study 3: Usability Evaluation in Daily Tasks}
\label{sec:study3}
One goal of this paper is to develop a practical user-defined mouth microgesture set and to understand which and how gestures would be used for real-life daily tasks. Beyond evaluating the physical and mental demand of these gestures alone, the context and environment of a user also has a significant impact on whether or not a user uses a gesture to perform an action.
In this section, we describe the design and insights from a third study to map the 14 preferred gestures from Study 2 to tasks in commonly used smartphone operations. 

\subsection{Task List and Gestures}
We first establish a list of applications and their relevant tasks that the microgestures can accomplish. We choose three types of applications that we believe represents a substantial range of applications people commonly use: audio, video, and textual (see Table \ref{tab:task_list}). These types indicate what users are mainly interacting with and what form of feedback they are receiving. For each type, we choose two applications with 1-3 common tasks. These tasks fall into three categories: (1) some are simple toggle inputs, (2) some take on a small discrete set of inputs, and (3) others have a continuous input. Certain operations can be controlled in both a discrete and continuous manner, as marked in Table~\ref{tab:task_list}.

The set of mouth gestures participants could choose to map to tasks was taken from the 14 preferred gestures from Study 2 that were the least physically and mentally demanding of their category. Because the one-bit gestures are actually two symmetrical zero-bit gestures, we separated them into distinct gestures for the purpose of this study. Therefore, participants had a set of 20 mouth gestures to choose from.  

\begin{figure*}[]
    \centering
    \includegraphics[width=0.8\textwidth]{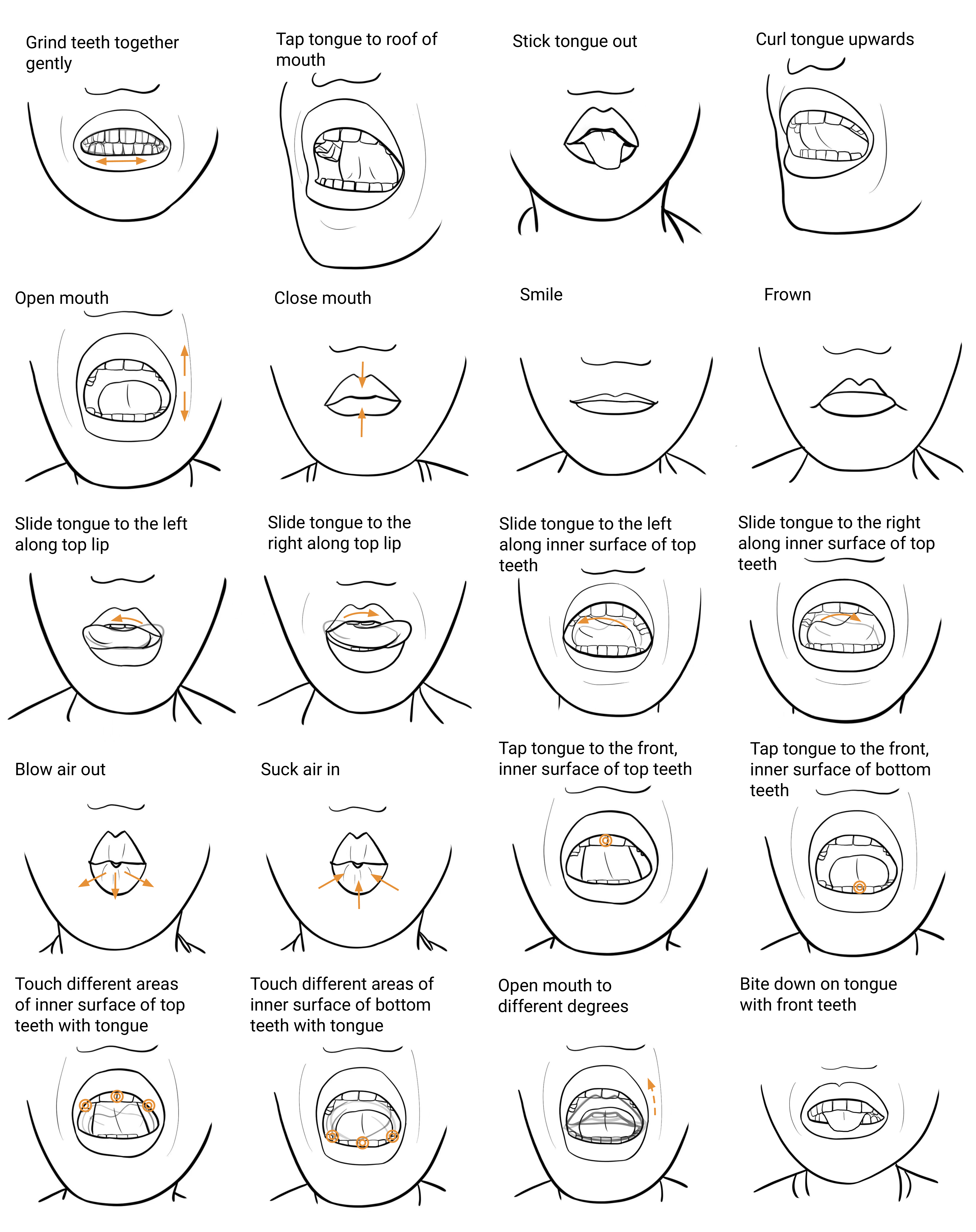}
  \caption{The filtered microgesture subset used for Study 3. The mouth is drawn slightly exaggerated and more open to make viewing the inner mouth more visible. These drawings are the authors' interpretation of the microgestures. Participants in the studies were not involved in their creation; they proposed or were exposed to descriptions of the gestures.}
  \label{fig:study3_drawings}
\end{figure*}

\subsection{Procedure}

We designed and conducted a remote within-group Wizard-of-Oz study to simulate an environment where participants would be using mouth gestures to control their smartphone.
Due to their ubiquitous nature, smartphones are often used in mobile and multitasking scenarios.
To consider the effects of these scenarios on the user's ability and perception of our gestures, our evaluation also included two contexts: the user is sitting down (inactive) or walking around (active). We recruited participants from those who took part in Study 1, and of the original 16, 11 agreed to return for this study. Participants individually met with one of the researchers over a remote video call using their mobile phone. They were also required to have a second device (e.g. laptop, tablet) to use. 

At the start of the study, participants were given a document with a list of descriptions of the 20 gestures. Participants then took a few minutes to practice performing each gesture and to ask the researcher any clarifying questions. They could then reference this list for the rest of the study on their secondary device. On the phone that they were using to join the video call, participants would see a shared screen of a mobile phone that the researcher could control. This setup helped the participant to feel like they were actually using a phone while also allowing the researcher to manipulate the visual feedback seen by the participants. 
\begin{table}
\centering
\small
\begin{tabular}{cp{150pt}c}
\specialrule{1pt}{1pt}{1pt}
\hline
\textbf{ID} & \textbf{Gesture Description} & \textbf{Category} \\
\hline
1  &  Grind teeth together gently & zero-bit \\
2  &  Tap tongue to roof of mouth & zero-bit \\
3  &  Stick tongue out & zero-bit \\
4  & Curl tongue upwards & zero-bit \\
5  & Bite down on tongue with  front teeth & zero-bit \\
6  &  Open mouth & one of one-bit  \\
7  &  Close mouth & one of one-bit  \\
8  &  Smile & one of one-bit  \\
9  &  Frown & one of one-bit  \\
10 &  Slide tongue to the left  along top lip & one of one-bit  \\
11 &  Slide tongue to the right along top lip & one of one-bit  \\
12 &  Slide tongue to the left along inner surface of top teeth & one of one-bit  \\
13 &  Slide tongue to the right along inner surface of top teeth & one of one-bit  \\
14 &  Blow air out & one of one-bit  \\
15 &  Suck air in & one of one-bit  \\
16 &  Tap tongue to the front, inner surface of top teeth & one of one-bit  \\
17 &  Tap tongue to the front, inner surface of bottom teeth & one of one-bit  \\
18 &  Touch different areas of inner surface of top teeth with tongue & multi-bit\\
19 &  Touch different areas of inner surface of bottom teeth with tongue & multi-bit\\
20 &  Open mouth to different degrees & multi-bit \\
\hline
\specialrule{1pt}{1pt}{1pt}
\end{tabular}
\caption{Gesture List}
\label{tab:gesture_list}
\end{table}

\begin{table}
\tabcolsep4pt\small\begin{tabular}{cccccc}
\specialrule{1pt}{1pt}{1pt}
\hline
\textbf{Type}            & \textbf{App}                & \textbf{Operation}  & \textbf{Property}  & \textbf{Cont-S} & \textbf{Cont-W} \\
\hline
\multirow{6}{*}{Audio}   & \multirow{3}{*}{Music}      & Vol Up/Down    & Disc/Cont & 12\&13 & 12\&13 \\
                         &                             & Start/Pause    & Toggle    & 2 & 16\\
                         &                             & Next/Pre Song  & Disc      & 19 & 19\\
\cline{2-6}
                         & \multirow{3}{*}{\shortstack{Phone\\Call}} & Pick Up        & Toggle    & 2 & 5\\
                         &                             & Mute/Unmute    & Toggle    & 4 & 4\\
                         &                             & Vol Up/Down    & Disc/Cont & 12\&13 & 12\&13\\
\hline
\multirow{5}{*}{Video}   & \multirow{2}{*}{Camera}     & Take Pic       & Toggle    & 5 & 5\\
                         &                             & Zoom In/Out    & Disc/Cont & 12\&13 & 12\&13 \\
\cline{2-6}
                         & \multirow{3}{*}{\shortstack{Video\\Call}} & Camera On/Off  & Toggle    & 5 & 2\\
                         &                             & Mute/Unmute    & Toggle    & 2 & 4\\
                         &                             & Switch Cam     & Toggle    & 1 & 5 \\
\hline
\multirow{3}{*}{Textual} & \multirow{2}{*}{Reading}   & Scroll Up/Down & Disc/Cont & 16\&17 & 16\&17\\
                         &                             & Inc/Dec Font   & Disc      & 14\&15 & 14\&15\\
\cline{2-6}
                         & Timer                       & Start/Stop     & Toggle    & 2 & 16\\
\hline
\specialrule{1pt}{1pt}{1pt}
\label{tab:tasks_agreements}
\end{tabular}
\caption[]{Task List\\
Table~\ref{tab:gesture_list} summarizes the 20 gestures obtained from the results of Study 2. The Category column indicates the property of each gesture.
Note that from No.6-17, we split the one-bit gestures into two separate zero-bit gestures, since each of them can be used independently. For example, No.10\&11 are a pair in Study 2 but can be chosen separately in Study 3.
Table~\ref{tab:task_list} lists out all operations and the property of each operation. Disc/Cont means it can either be controlled discretely or continuously (\eg volume adjustment).
The last two columns are the most commonly selected gestures for each tasks under the sitting(S)/walking(W) context after solving conflicts.
}
\label{tab:task_list}
\end{table}

The study continued as follows: the researcher verbally prompted the participant with a scenario that they were using a specific phone application to complete a task. The researcher would display the application and the before and after effect of an operation to the participant's phone screen. The participant was given some time to look at the list of microgestures and choose one they would want to use. Once they had decided, they would declare to the researcher that they had made their choice and then immediately perform their chosen gesture. After hearing their declaration, the researcher completed the operation on the phone so the participant could see the resulting visual feedback. The participant then informed the researcher which gesture from the list they would prefer to use. After each scenario, the researcher asked three questions to the participant: (1) whether they preferred the gesture they used over an alternative non-gesture interaction for the task (see Table \ref{tab:study3:baselines}), (2) whether the gesture was easy to perform, (3) whether they thought the microgesture was socially acceptable. Participants answered with a rating from a 7-point Likert scale. This exchange would repeat for all of the tasks for both sitting and walking contexts. All tasks when the participant was sitting were completed together and likewise when they were walking. The sitting and walking contexts were counterbalanced across users, and the applications participants saw were ordered with a balanced Latin square (n=6). In total, each participant completed 28 tasks. 
\begin{table*}[t]

\small
\begin{tabular}{ccl}
\specialrule{1pt}{1pt}{1pt}
\hline
\textbf{App}  & \textbf{Operation}   & \textbf{Baseline Interaction}  \\
\hline
\multirow{3}{*}{Music}   & Vol Up/Down  & Using plus/minus (or equivalent) button controls on earphones \\
                        & Start/Pause  & Using button controls on earphones \\
                        & Next/Pre Song  & Triple/long press (or equivalent) button controls on earphones \\
\hline
\multirow{3}{*}{Phone Call}  & Pick Up   & Using button controls on earphones \\
                             & Mute/Unmute  & Tapping mute button on phone touchscreen \\
                             & Vol Up/Down  & Triple/long press (or equivalent) button controls on earphones\\
\hline
\multirow{2}{*}{Camera}   & Take Pic   & Pressing volumn controls on side of phone \\
                             & Zoom In/Out  & Using two fingers to pinch on the phone touchscreen\\

\hline
\multirow{3}{*}{Video Call}   & Camera On/Off   & Tapping button on the phone touchscreen \\
                             & Mute/Unmute  & Tapping button on the phone touchscreen \\
                             & Switch Camera  & Tapping button on the phone touchscreen\\
                             
\hline
\multirow{2}{*}{Reading}   & Scroll Up/Down   & Using finger to swipe on phone touchscreen \\
                             & Inc/Dec Font  & Using two fingers to pinch on the phone touchscreen \\
\hline
            Timer   &  Start/Stop   & Using a voice command \\

\hline
\specialrule{1pt}{1pt}{1pt}
\end{tabular}
\caption{Participants rated their preference of the mouth microgesture compared to an alternative, existing interaction technique. When possible, non-touchscreen interactions were used as baselines to make a more fair comparison, because mouth microgestures are hands- and eyes-free. }
\label{tab:study3:baselines}

\end{table*}

When choosing the gesture to map to a task, participants were not allowed to use a gesture for more than one task in a specific application. However, across different applications, participants could pick the same gesture. The exceptions to this are the multi-bit gestures 18, 19, and 20. Since they are more open-ended and execution depends on the user, we let participants reuse the same multi-bit gesture within an application. When notifying the researcher which gesture they used for a task, participants described the differences in the performance if they used a multi-bit gesture multiple times. 

\subsection{Results}
\label{sub:study3:results}
Because the one-bit gestures were divided and mixed in with zero-bit gestures, the meanings of zero- and one-bit become more fluid, conforming to however the participants used them. 
For the rest of this analysis, we define \textit{singular} and \textit{paired} gestures, which are functionally equivalent to our previous definitions of zero-bit and one-bit, respectively. 
When we specify zero- or one-bit microgestures, we refer to their original sense in Study 2. 
A singular gesture can be a zero-bit gesture or one side of a one-bit gesture, and a paired gesture can either be two sides of any one-bit gesture, or a combination of two zero-bit gestures. 
\subsubsection{Most Selected Gestures}
\label{subsub:study3:results:most_selected}
\begin{figure*}[]
    \centering
    \includegraphics[width=0.9\textwidth]{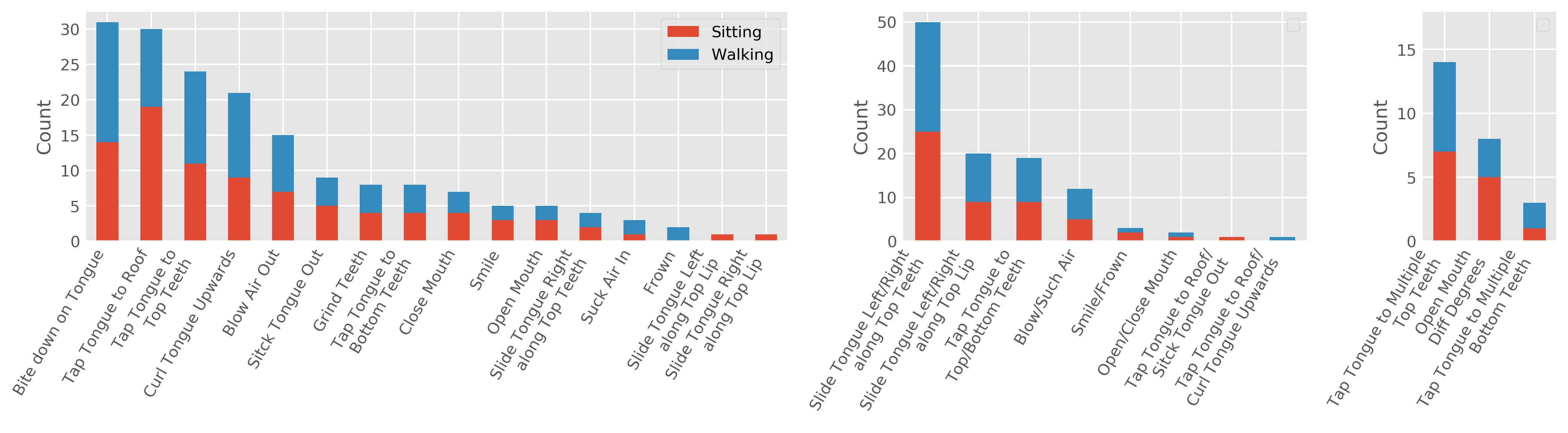}
  \caption{Selection counts of each gesture among the 14 audio, video, and textual tasks under both sitting and walking conditions. The left figure shows the stacked bar plot of the counts for all zero-bit gestures (No.1-5), together with the cases where one of the one-bit paired gestures is used alone (No.6-17). The middle figure shows the plot for all one-bit gestures, including the cases where two zero-bit non-pair gestures are used together. And the right figure shows the plot for all multi-bit gestures (No.18-20).}
  \label{fig:study3_selection_counts}
\end{figure*}

Participants only picked singular gestures for \textit{toggle} operations, including zero-bit gestures, G1-G5, and one side of one-bit gestures, G6-G17 (see Table~\ref{tab:gesture_list}).
For \textit{disc/cont} operations, most participants either selected paired gestures or multi-bit gestures. Of the paired gestures, all followed the original one-bit definitions with only two exception: P4 picked gesture 2 and 3 for the music-volume up/down task in the sitting context, and P10 picked gesture 2 and 4 for the music-next/previous song task in the walking context.
We summarized the number of times gestures in each category (singular, paired, multi-bit) were selected in Figure~\ref{fig:study3_selection_counts}.

There were a few observations.
First, among singular gestures (the left of Figure~\ref{fig:study3_selection_counts}), the top 5 gestures - G5, G2, G16, G4, G14 - were consistent between the sitting and the walking context, and account for 70\% of the total singular gesture count.
It was interesting to see that two of the five gestures were from one-bit gestures (G16 and G14).
However, these are the only two with top rankings.
Overall, zero-bit gestures were more preferred than one side of one-bit gesture for toggle operations. The top 7 gestures included all five zero-bit gestures and the average times that a zero-bit gesture was selected was four times of that of a one side one-bit gesture (21.4 \vs 5.3).

Second, the middle of Figure~\ref{fig:study3_selection_counts} indicates that the paired gestures G12~\&~G13 (\textit{slide tongue to the left~\&~right along inner surface of top teeth}) were significantly more preferred than the other paired gestures for discrete or continuous control in both sitting and walking contexts (generalized linear mixed models (GLMMs) with binomial family~\cite{mcculloch2005generalized}, $p$s $< 0.05$). This single pair accounts for 46\% of cases. Moreover, the top four pairs - G12~\&~G13, G10~\&~G11, G16~\&~G17, G14~\&~G15 - accounted for 94\% of the cases, indicating that the other pairs (\eg G6~\&~G7 Open/Close mouth, and G8~\&~G9 Smile/Frown) were rarely picked by users.

Third, among the three multi-bit gestures (only selected for dist/cont operations), the most selected was gesture 18 (56\% of the cases), followed by gesture 20 (32\%) and 19 (12\%).
Although gesture 18 and 19 were similar and symmetric gestures, with the only distinction on touching the top versus bottom part of the teeth, it was interesting to find that their preference differs significantly, which was different from the rankings in Study 2 (Table~\ref{tab:physicalmentalscores}).
We discuss this observation further in Section~\ref{subsub:study3:results:symmetry}.

\subsubsection{Symmetry and Asymmetry of One-bit Gestures}
\label{subsub:study3:results:symmetry}
In Study 2,
participants rated one-bit gestures, disregarding their preference between the two sides in the case of sided gestures.
In this study, the selections made for the toggle operation provided the opportunity to compare preference for the two sides.

As we noted in Section~\ref{subsub:study3:results:most_selected}, gestures G14 and G16 were among the top 5 commonly selected singular gesture. However, the gestures on the other side (G17 and G15) were significantly less preferred ($p$s $< 0.05$), whose selected times were only 36\%~/~31\% and 14\%~/~25\% of their respective pairs in the sitting/walking context.
Moreover, even for the multi-bit gestures G18~\&~G19, touching areas on the bottom teeth were only selected 14\%~\&~28\% of the time for touching areas on the top/bottom teeth in the sitting/walking context.
These results show asymmetry of one-bit gestures: using the top teeth was more preferred than using the bottom teeth for interaction, and blowing air out was more preferred than sucking air in. 
While this contrasts with the rankings from Study 2 that place gesture 19 above 18, we believe that this asymmetry relates to our observations in Section \ref{subsub:study2:results} about sufficient haptic feedback and proprioception. The tongue naturally resides near the bottom teeth which does indeed make it less physically demanding to reach with the tongue. However, it lacks the purposeful feedback that touching the top teeth has when users are actively moving their tongue to the slightly higher position. 

In contrast, the preference of left and right gestures was more symmetric. They were rarely selected alone for toggle operations (2/2 times for G13, 0/1 for both G10 and G11 in the sitting/walking context). In addition, the results of GLMMs did not show significance between the two sides of G12~\&~G13 and G10~\&~G11  ($p>0.05$).

\subsubsection{Users' Preference under Different Contexts}
\label{subsub:study3:results:agreement}
We compared the most common gestures picked by participants in each task in the two contexts.
Nine of the fourteen tasks had either the same or overlapping most selected gestures (in some tasks, more than one gesture tied for the maximum selections).
We performed Fisher's exact tests~\cite{upton1992fisher} on each task. The results did not indicate any significance between the sitting and the walking context ($p>0.05$).

\begin{figure*}[]
    \centering
    \includegraphics[width=0.6\textwidth]{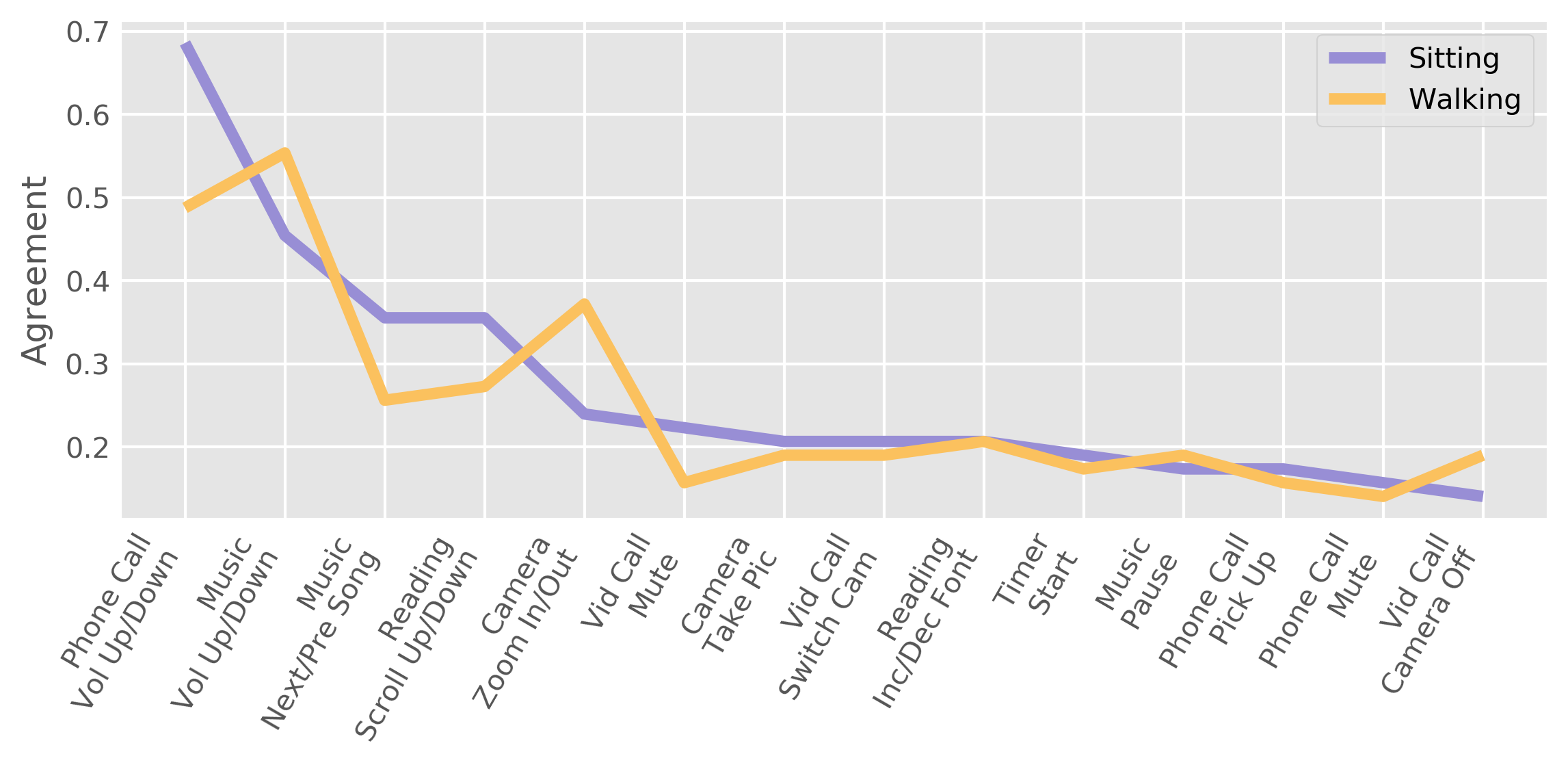}
  \caption{Gesture selection agreement among participants for the 14 audio, video and textual operations.}
  \label{fig:study3_agreement}
\end{figure*}

We further investigated the gesture selection agreement for each task. The agreement score $A_o$ for an operation $o$ was calculated by
$$A_o = \sum_{P_i \subseteq P_o}\left(\frac{|P_i|}{|P_o|}\right)^2$$
where $P_o$ is the set of selected gestures for the operation, and $P_i$ is a subset of identical gestures from $P_o$~\cite{wobbrock-user-defined,wobbrock2005maximizing}.
Figure~\ref{fig:study3_agreement} visualizes the scores of each task under the two contexts.

Interestingly, we found that the top five operations were all disc/cont operations (six in total), and that the score for most toggle operations were close to 0.2. This indicates that users' preferences were less diversified for disc/cont operations.

The volume adjustment tasks in both the phone call and music player applications had the highest agreement scores, no matter if users were in the sitting or walking context. The most commonly selected gestures for volume adjustment were G12~\&~G13 (\textit{slide tongue to the left/right along inner surface of top teeth}).
Therefore, we mapped these two gestures for volume up/down in both the phone call and music player applications.
Following the same procedure, we continued to find the most commonly selected gestures for all tasks one after another, in two contexts separately.
In the same application, if two operations had the same gestures, such a conflict was solved by having the larger group win the gesture.
Our gesture mapping results are summarized in the last two columns in Table~\ref{tab:task_list}.
The finalized gesture set covers 63.6\% of the agreement for the sitting context and 62.6\% for the walking context.

\subsubsection{Comparison to baseline interactions}
One aspect of mouth microgestures we were interested in is not just users' preference between the gestures but also in general, their preference over standard interaction techniques for specific tasks. For each task, we took the mean of all participants' ratings for comparing to a baseline. The mean aggregated score for a task was 4.97 (stdev=0.53) when participants were sitting down, and 5.04 (stdev=0.44) when walking around. 
This suggests that the tasks are, for most users, conducive to using mouth microgestures as an interaction.
Some of the tasks that had lower mean scores in both contexts and did not clearly favor gestures as much were \textit{camera on/off} ($\Bar{x}=4.0$ for sitting, $\Bar{x}=4.36$ for walking) for the \textit{video call} application and \textit{increase/decrease font} ($\Bar{x}=4.0$ for sitting, $\Bar{x}=4.82$ for walking) for the \textit{reading} application.

\subsubsection{Performance Workload}
\label{subsub:study3:results:easy}
To validate our rankings from Study 2, we analyze the results of participants' Likert scale answers to if they thought the gesture they chose was easy to perform. For each gesture, we calculated the overall mean rating over all participants and then calculated the mean and standard deviation of the aggregated ratings. Overall, we found that participants did indeed regard the subset of gestures we presented as easy to perform in both contexts of sitting and walking (mean=5.82, stdev=0.54 for sitting; mean=6.04, stdev=0.37, for walking), supporting the results on physical and mental demand from the previous study. In the sitting context, we notice that the lowest mean ratings belonged to the G1\&G18.

\subsubsection{Social Acceptability}

We take a similar approach as in Section \ref{subsub:study3:results:easy} and again aggregate the ratings for each microgesture by taking the mean. The results show that participants generally thought the microgestures were socially acceptable (mean=5.42, stdev=1.36 for sitting; mean=5.42, stdev=1.42, for walking), but there are a few microgestures that stand out to be less socially acceptable, as indicated by the increased aggregated variance in scores. Gestures G3 ($\Bar{x}=2.3$), G6 ($\Bar{x}=3.5$), and G20 ($\Bar{x}=3.4$) had the lowest ratings when the participants were sitting down, and gestures G3 ($\Bar{x}=1.75$) and G20 ($\Bar{x}=3.33$) were also rated low in the walking context. A common feature of these gestures was that they are visually noticeable by a third party. It is interesting though that participants still selected these gestures rather often, as shown in Figure \ref{fig:study3_selection_counts}.
\section{Discussion}
\label{sec:discussion}
In this section, we describe how the results of the user studies have implications on the design of mouth microgestures with end-users in mind. 
\subsection{Design Implications}

\subsubsection{Design Guidelines for Mouth Microgestures}
We deduce several general design guidelines for novel mouth microgestures and summarize them here. 

First, we found that \textbf{short, direct actions as gestures were preferable}. Microgestures that were intricate and required compound or sequential motions were not regarded as highly as simpler ones.

Closely tied to the previous is that \textbf{having fewer moving organs involved was better}. Gestures requiring manipulation of multiple parts of the mouth in conjunction with each other were often rated poorly by users.  
We speculate that this could be because the mouth is still an unfamiliar mode of interaction, making complex actions unfavorable, or that the speed of gesture execution could be important when using the mouth. 

The next guideline we report is that \textbf{natural mouth movements are good for intuition but not necessarily preferred as a gesture}. Natural motions, like smiling, may produce gestures that are better for learnability and memorability but these such actions were chosen infrequently. It is possible that for end users, the similarity to everyday actions makes them poor choices for gestures which are meant to be deliberate. 

Next, we note that \textbf{location and direction have strong meanings in the microgesture}; details like whether an action is toward the left/right or upward/downward can affect a user's mental model of a microgesture and what they expect it to do. 

Lastly, we suggest that \textbf{proprioception is important for performing eyes-free gestures}. This refers to movements that provide haptic feedback for itself, such as tapping the tongue against the roof of the mouth and feeling the roof of the mouth with the tongue. Because mouth microgestures are eyes-free, users need to rely on another form of feedback during execution to feel confident that the microgesture was correct.

\subsubsection{Implications for Mouth Microgesture Systems}
When viewing the distribution of proposed gestures from Study 1, a considerable number of them make up the group using the \emph{tongue} as the primary organ. Taking this into account, it is advisable to avoid obstructing the range of motion of the tongue when developing the sensing system for a mouth microgesture interface. On a similar note, for the other axis of our taxonomy, most microgestures fall in the \emph{state} category. This means that these gestures rely on position or location within/around the mouth. Considering how many there are in this group, fine-grained localization of mouth parts may be a useful feature to pursue when implementing a technical system.

The taxonomy we propose in this paper could also guide how to organize one's mouth microgesture set. For instance, certain functionality in the user interface may be fairly different or unrelated (e.g. volume control vs. navigation) and the mapped gestures similarly should be distinct. Following the defined groupings of mouth microgestures, a gesture designer could ensure that gestures for each functionality use different categories, like separate primary organs or different modalities.

\subsubsection{Factors Influencing Users' Preference of Gestures}
Many interfaces for different applications have similar functional widgets, like having some directional action or a continuous input slider. Gesture reusability across applications has much value as it can help with discovering and remembering mouth microgestures, especially since they are currently a novel interaction. As seen in Section \ref{subsub:study3:results:most_selected}, the microgestures \textit{slide tongue to the left/right along inner surface of top teeth} were often selected for different tasks in various applications.

Even though mouth microgestures do not rely on visual feedback to be useful, user preference may still be influenced by past experiences with touchscreen interfaces. 
Mouth microgestures that can be drawn from analogous existing gesture interfaces (i.e. tap, swipe) were found to be most popular in our studies, likely due to existing familiarity.
Tongue microgestures share a similar mental model with that of finger gestures on a touchscreen, and since interfaces are often still designed around the touchscreen experience, the spatial reasoning of using a finger may carry over to using mouth microgestures. 
In study 3, we also noticed some participants chose microgestures associated with rightward motion, like \textit{slide tongue to the right along inner surface of front teeth}, for the task of answering a phone call. Upon closer examination, we realized that this may have been due to the user interface used to display the \textit{referent} which indicated swiping to the right to pick up the call and left to hang up. We found that gestures were closely tied with operations, independently verifying results from Wobbrock et al’s findings on user-defined gestures in surface computing~\cite{wobbrock-user-defined}.

Metaphors associated with everyday actions with the mouth, like those related to communication, also may play a role in the design of a mouth microgesture set. P6 from Study 3 chose the \textit{close mouth} for three of four "mute" tasks for the phone or video call applications. They described how it seemed the most intuitive if they wanted to stop the sound input. This comment suggests to us that natural movements of the mouth carry meaning that can be applied to interaction; some of the user-defined microgestures like \textit{clear throat} or \textit{make a short 'snoring' sound} may have intuitive purposes as an interaction.

%


\subsection{Technological Context}

From the results of our third study, we derived a practical set of microgestures that can be mapped to common smartphone interactions. 
While we planned our studies to explore the design space to discover the ideal user-designed gestures, free of constraints, it is meaningful to relate our findings within the current technological landscape in regards to sensing capabilities. 
In our related work section, we reviewed prior work showing that many facial movements can be detected with head-mounted systems and commodity sensors. 
We expect a variety of these existing sensing techniques to be capable of differentiating between our proposed gestures. 
Many of the final selected gestures involve tongue movements to different areas of the mouth which have been shown to be possible to detect in~\cite{cheng2014tip},~\cite{Nguyen-TYTH}, and~\cite{goel2015tongue}.
We envision that recent work using sensors around the ear including acoustic sensing~\cite{Amesaka2019}, electric field sensing~\cite{Matthies2017} and motion sensor~\cite{10.1145/3290607.3312925}, are promising avenues to making mouth microgestures more adoptable.
Some of the microgestures may be subtle enough to pose a challenge to detect with a single sensing modality. We believe, however, that a sensor fusion approach can overcome these cases, and as recent wearables around the face are being equipped with more sensors, the issue of insufficient or unreliable data will be less problematic.
Our studies reveal important gesture characteristics and insights of what users expect for this new type of interface, and this knowledge can be helpful for providing direction when developing future interfaces and technical systems.

\subsection{Limitations}
The taxonomy we define was developed solely from the gathered gestures of Study 1. Since we did not supplement this set with the gestures used by past technical papers, there may be a loose connection between the insights in gestures design that we derive and any technological implications.

Because of the way we defined our taxonomies of mouth microgestures, participants in the user studies were only exposed to microgestures unique in their spatial design. Variations in their execution, many of which participants in the initial brainstorming session proposed, were not considered. Many microgestures could be performed two or three times as a new microgesture or one could use temporal variations of microgestures like performing them more quickly/slowly. These kinds of modifications may influence a user's choice of using a new microgesture for a task or using a microgesture variation.

The design of the Wizard-of-Oz study has a few drawbacks. In order to smoothly facilitate the study remotely, the primary mode in which \textit{referents} were administered to participants was visually through a smartphone, so application interfaces were constrained to those normally designed for a mobile phone. People's interactions with wearables with minimal screens or hearables may involve other forms of feedback that our study design does not effectively capture. Mouth microgestures have the advantage of being both hands- and eyes- free, so participants may not have fully experienced this feature. Also, the contexts we tested only capture a limited range of the wide spectrum of possible user activities. There may be other common daily scenarios that could influence a users' choice of microgesture. Multiple participants commented for the reading task that if their hands or fingers were dirty or occupied, like when cooking, then they would have rated their preference of the microgesture over the baseline touch interaction much higher.

\section{Conclusion}
\label{sec:con}
We explore the design space of mouth-based microgestures and analyze users' perception of them as an interaction technique to accomplish routine tasks with their personal devices. 
From an original set of 86 collected user-defined gestures, we present taxonomies to characterize how mouth microgestures can be formed and applied. 
We present a functional set of 20
mouth microgestures, determined by user preference, that can be applied to tasks of common software applications.
The insights we've learned on user behavior of mouth microgestures should help future interaction designers of wearables and hearables develop intuitive, usable mouth microgestures.

\begin{acks}
This work is supported by the National Key R\&D Program of China under Grant No. 2019YFF0303300, the Natural Science Foundation of China under Grant No. 62002198. Our work is also supported by the Beijing Key Lab of Networked Multimedia, Undergraduate / Graduate Education Innovation Grants, Tsinghua University. We would like to thank all participants for their time and effort.
\end{acks}

\balance
\bibliographystyle{ACM-Reference-Format}
\bibliography{ref}

\end{document}